\documentclass[options]{JHEP3}   

\JHEPspecialurl{http://jhep.sissa.it/JOURNAL/JHEP3.tar.gz}

\usepackage{epsfig,multicol}

\def\be{\begin{eqnarray}}
\def\ee{\end{eqnarray}}
\newcommand\fverb{\setbox\pippobox=\hbox\bgroup\verb}
\newcommand\fverbdo{\egroup\medskip\noindent%
            \fbox{\unhbox\pippobox}\ }
\newcommand\fverbit{\egroup\item[\fbox{\unhbox\pippobox}]}
\newbox\pippobox

\title{
\begin{flushright}
\normalsize{ FERMILAB-Pub-06/277-T\\UV-06-0818}
\end{flushright}
Colliders as a simultaneous probe of supersymmetric dark
matter and Terascale cosmology}

\author{       {Gabriela Barenboim}\\
\normalsize\emph{Departament de F\'isica Te\`orica, Universitat de
Val\`encia}\\ 
\emph{Carrer Dr. Moliner 50, E-46100 Burjassot (Val\`encia), Spain}\\
Email: \email{gabriela.barenboim@uv.es}\\}
\author{\textbf{Joseph D. Lykken}\\
\normalsize\emph{Fermi National Accelerator Laboratory}\\
\emph{P.O. Box 500, Batavia, IL 60510, USA}\\
Email: \email{lykken@fnal.gov}}

\abstract{Terascale supersymmetry has the potential to
provide a natural explanation of the dominant dark matter component of the standard
$\Lambda$CDM cosmology. However once we impose the constraints on minimal
supersymmetry parameters from current particle physics data, a satisfactory
dark matter abundance is no longer {\it prima facie} natural. This Neutralino
Tuning Problem could be a hint of
nonstandard cosmology during and/or after the Terascale era. To quantify this
possibility, we introduce an alternative cosmological benchmark based upon
a simple model of quintessential inflation. This benchmark has no free parameters,
so for a given supersymmetry model it allows an unambiguous prediction
of the dark matter relic density. As a example, we scan over the parameter
space of the CMSSM, comparing the neutralino relic density predictions with
the bounds from WMAP. We find that the WMAP--allowed regions of the CMSSM
are an order of magnitude larger if we use the alternative cosmological benchmark,
as opposed to  $\Lambda$CDM. Initial results from the CERN Large Hadron Collider
will distinguish between the two allowed regions.}


\begin{document}


\section{Introduction}
One of the most appealing features of Terascale supersymmetry \cite{Chung:2003fi}
is the potential to
provide a natural explanation of the dominant dark matter component of the standard
$\Lambda$CDM cosmology.  This explanation is driven by particle physics
motivations from data which are completely unrelated to the astrophysical data
that motivate $\Lambda$CDM. In most supersymmetry (SUSY) models the dark matter
candidate is the neutralino, although there are other interesting possibilities. 
The mass and annihilation cross section of the neutralino are determined by
the SUSY model parameters. Adapted to a $\Lambda$CDM cosmological history,
the neutralino becomes a cold thermal relic, freezing out during a radiation-dominated
era at a temperature equal to approximately 1/20th of its mass. SUSY model
parameters thus determine the current neutralino abundance, which is usually
expressed as a fraction, $\Omega_{\chi}$, of the critical density.

If we assume that neutralinos constitute nearly all of the non-baryonic matter at
late times, then their abundance is quite constrained by astrophysical data.
For example, combining the WMAP three year data with Sloan Digital Sky
Survey data on large scale structure, one obtains \cite{Spergel:2006hy}
the (naive) 2-sigma limits:
\be\label{eqn:wmapbounds}
0.095 < \Omega_{\chi}h^2 < 0.122 \; .
\ee
Thus the neutralino relic abundance is determined with $\sim 10$\%\ accuracy.

Suppose for simplicity that
we consider some subclass of Terascale SUSY models with the neutralino
as the stable lightest supersymmetric partner (LSP). The most popular example
is the CMSSM, inspired by minimal supergravity \cite{Barbieri:1982eh}-\cite{Nath:1983aw},
where the SUSY parameter
set is written $m_0$, $m_{1/2}$, $A$, tan$\,\beta$, and sign($\mu$).  Imposing
very rough considerations of naturalness, we can require $m_0$, $m_{1/2}$
and $|A|$ to be less than, say, 3 TeV.
If we assume
$\Lambda$CDM cosmology we can scan over this parameter space,
placing a mark at each point which would predict a relic density satisfying
the ``WMAP'' constraints  (\ref{eqn:wmapbounds}). 

\begin{figure}
\centerline{\epsfxsize 3.0 truein \epsfbox {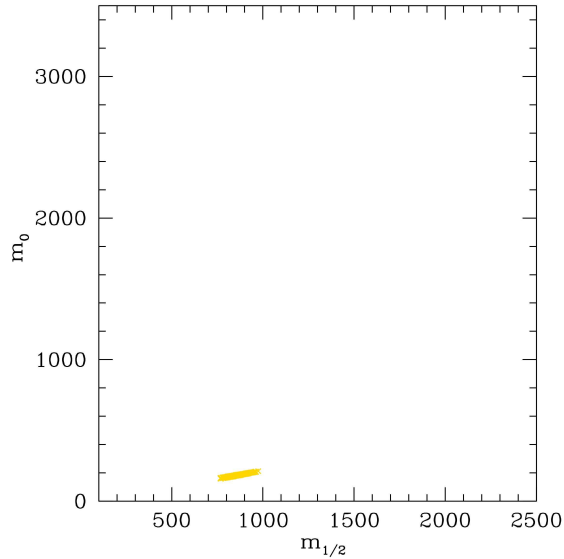}}
\caption{\label{fig:sm5hfs} The WMAP--allowed points
for the CMSSM, with $A=0$, $\mu >0$, and tan$\,\beta =5$.
Points are shown only if the corresponding Higgs mass is $\ge 114$ GeV.
The scan was performed in 2 GeV increments.
\hbox to170pt{}}
\end{figure}

The results look like Figures \ref{fig:sm5hfs} and \ref{fig:sm30fs}, where we
have chosen $A=0$, $\mu >0$, tan$\,\beta = 5$ or 30, and we only plot points which also
satisfy the existing experimental bounds on the superpartner particles and the Higgs.
The ``WMAP--allowed'' regions are a very small fraction of the total parameter
space. More significantly, once we impose experimental bounds the WMAP--allowed points 
are not generic; for a Bino-like
neutralino generic points predict a relic density that is much too large, while
for a Higgsino-like or Wino-like neutralino generic points predict a relic density that
is much too small. Getting a WMAP--allowed relic density thus requires tuning to 
regions where various conspiracies take place, either enhancing the neutralino
annihilation (or co--annihilation) cross section, or balancing the 
Bino-Higgsino-Wino content of the LSP.
These conspiracies can be quantified by defining simple
sensitivity measures of the predicted relic density to small variations
in relevant SUSY parameters \cite{Arkani-Hamed:2006mb}.
Since conspiracy is the opposite of naturalness, 
such an analysis \cite{Baer:2002gm}-\cite{Mambrini:2005cp}
casts doubt
upon the CMSSM, and perhaps Terascale SUSY in general, as a natural explanation
of dark matter.

This ``Neutralino Tuning Problem'' is similar to the much more famous ``Little Hierarchy
Problem'' \cite{Barbieri:2000gf,Birkedal:2004xi} of the MSSM. 
While conceptually distinct, these problems are related
by the fact that the superpartner and Higgs mass bounds from data eliminate
what would otherwise be a generic WMAP--allowed region of the CMSSM known as
the ``bulk'' region. The bulk region is generic because its only distinguishing
feature is that the superpartners are light, as favored by naturalness applied to
electroweak symmetry breaking. Thus at least in the CMSSM the two naturalness
problems are coupled.

In this paper we revisit the important question of whether we can quantitatively evaluate
the claim that Terascale supersymmetry provides a natural explanation of
dark matter.
This issue has not been fully resolved in the literature,
due to a number of difficulties which will we review in the next section.
We will show that a physically meaningful approach to this question is to
compare the robustness of the connection
between SUSY and dark matter under alternative well-motivated cosmological
scenarios.

There is ample motivation to vary our cosmological assumptions.
Despite the dramatic success of the standard $\Lambda$CDM model
in confronting a wide variety of astrophysical data, this same data is largely
silent on the cosmological history of the universe before big bang
nucleosynthesis (BBN), {\it i.e.}, at temperatures above about 4 MeV.
WMAP data is consistent with an epoch of primordial inflation, but we are
very far from being able to flesh out these hints into a data-driven model
which is sufficiently robust and detailed to make definitive statements
about dark matter. 

A difficulty in exploring cosmological alternatives
is that, because Terascale cosmology\footnote{In this
phrase `Terascale' refers to the physics responsible for the WIMP dark matter,
not the freeze--out temperature.}
is almost a black box, 
it is easy to introduce new parameters which are relevant but
unconstrained, modifying
the neutralino relic density predictions by orders of magnitude. For example, we can dilute
the neutralino density by adding entropy to the radiation bath some
time after neutralino freeze--out but before BBN. Alternatively we can increase the
predicted neutralino abundance by producing them non-thermally.  
Our solution is to look at alternative cosmological scenarios which are
reasonably motivated, make definite predictions for dark matter, and are
sufficiently simple and constrained that they have no relevant adjustable
parameters. The dark matter predictions of Terascale SUSY can then be compared
for these cosmological benchmark scenarios.

As a first alternative cosmological benchmark, we propose the Slinky model
of quintessential inflation described 
in \cite{Barenboim:2005np,Barenboim:2006rx}. This model has no
adjustable parameters once we require that the current radiation and
dark energy fractions take their WMAP--preferred values, and that the
universe is overwhelming radiation-dominated during the time of BBN.
Since the same inflaton is responsible for primordial inflation and for
dark energy, it is not surprising that the universe is not 
completely 
radiation-dominated at the time that neutralinos freeze out. By making the minimal
assumption that inflatons do not decay to neutralinos, we have a picture
in which thermally produced neutralinos are diluted by a predictable amount,
over and above the dilution of standard $\Lambda$CDM.
This additional dilution of the neutralino abundance can be expressed
(to an accuracy of a few percent) as a polynomial function of the
square root of the neutralino freeze--out temperature.

Thus for any SUSY model we can compute the predicted neutralino
relic abundance for two contrasting cosmological benchmarks,
$\Lambda$CDM and Slinky. The ratio of the WMAP--allowed regions
is a physically meaningful naturalness comparison. In our CMSSM scan,
we find that this ratio is always small. For
the Slinky benchmark the WMAP--allowed points are more generic, without
balanced mixings, mass degeneracies, or resonance-inducing mass relations.

These results give a concrete measure of the Neutralino Tuning Problem
in the CMSSM. At the same time, they define the beginnings of a straightforward program
to probe Terascale cosmology at colliders.

\section{The Neutralino Tuning Problem}

This problem is most easily observed in (but not limited to)
the CMSSM. We bound the parameter space by requiring $m_0$, $m_{1/2}$
and $|A|$ to be less than, say, 3 TeV. We remove regions in which the lightest
neutralino is not the LSP, or in which we do not get proper electroweak
symmetry breaking. We also remove regions which are in direct conflict
with experiment. For most of the MSSM parameter space, the most restrictive
experimental constraint is the direct lower bound on the lightest Higgs
mass from LEP. We will conservatively write this bound as $m_h \ge 114$ GeV.
We fix the mass of the top quark to the 2005 combined Tevatron average \cite{unknown:2005cc}
of 172.7 GeV; our results would differ only slightly using the newer combined
average \cite{Group:2006xn} of 171.4 GeV.

\begin{figure}
\centerline{\epsfxsize 3.0 truein \epsfbox {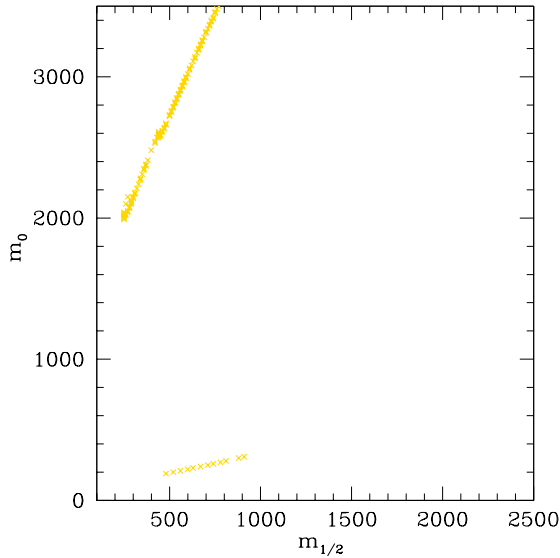}}
\caption{\label{fig:sm30fs} The WMAP--allowed points
for the CMSSM, with $A=0$, $\mu >0$, and tan$\,\beta =30$.
Points are shown only if the corresponding Higgs mass is $\ge 114$ GeV.
The scan was performed in 10 GeV increments, which misses a few of
the highest mass points in the co-annihilation region (lower right).
\hbox to170pt{}}
\end{figure}

\begin{figure}
\centerline{\epsfxsize 3.0 truein \epsfbox {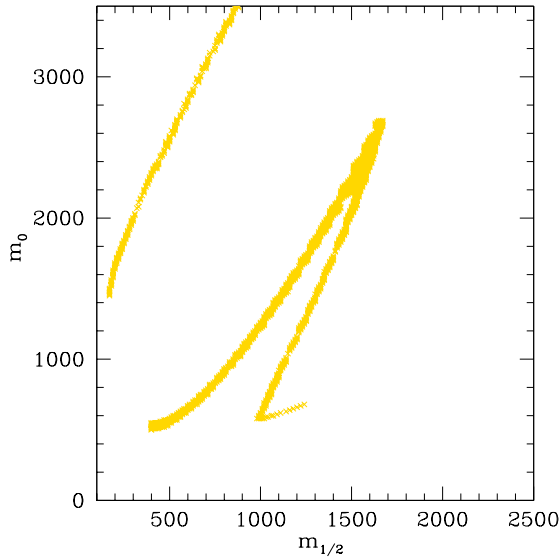}}
\caption{\label{fig:sm50fs} The WMAP--allowed points
for the CMSSM, with $A=0$, $\mu >0$, and tan$\,\beta =50$.
Points are shown only if the corresponding Higgs mass is $\ge 114$ GeV.
The scan was performed in 10 GeV increments.
\hbox to170pt{}}
\end{figure}

Assuming
$\Lambda$CDM cosmology we can scan over the remaining CMSSM parameter space,
placing a mark at each point where the computed neutralino relic density satisfies
the WMAP constraints (\ref{eqn:wmapbounds}). 
Representative results are seen by fixing $A=0$, $\mu >0$, and choosing either
tan$\,\beta = 5$ (Figure \ref{fig:sm5hfs}), tan$\,\beta = 30$ (Figure \ref{fig:sm30fs}) or
tan$\,\beta = 50$ (Figure \ref{fig:sm50fs}).
The figures are essentially identical to those in \cite{Djouadi:2006be}, except
that our allowed regions are smaller; this is because we use the new WMAP 2-sigma bounds,
while \cite{Djouadi:2006be} uses the older WMAP 99\% cl bounds, which are less
constraining.

Once we impose experimental bounds the WMAP--allowed points of the CMSSM
are not generic. For a Bino-like
neutralino generic points predict a relic density which is much too large, while
for a Higgsino-like or Wino-like neutralino generic points predict a relic density which 
is much too small. Getting a WMAP--allowed relic density thus requires tuning to 
regions where various conspiracies take place, either enhancing the neutralino
annihilation (or co--annihilation) cross section, or balancing the 
Bino-Higgsino-Wino content of the LSP.

In Figures \ref{fig:sm5hfs} and \ref{fig:sm30fs}, 
the lower slivers of WMAP--allowed points are
made possible by tuning the lightest stau to be nearly degenerate with
the lightest neutralino, producing co-annihilation at the time of
freeze--out. The upper sliver in Figure \ref{fig:sm30fs} is a ``focus point''
region \cite{Feng:2005ee}, 
where we tune to give the LSP a significant Higgsino component.
This allows neutralino pairs to annihilate more efficiently
through $t$-channel chargino exchange into dibosons. The focus point
region does not show up in Figure \ref{fig:sm5hfs} because it produces
only points with $m_h < 114$ GeV.
Apart from this, shifting tan$\,\beta$ from 5 to 30 does not make a qualitative difference.

For a window of very large tan$\,\beta$, which is itself a tuning,
we obtain results as in Figure \ref{fig:sm50fs}. Here the focus point
region has expanded, the co-annihilation sliver remains a sliver, and
a new region, the ``$A$--funnel'', has opened up. In this region we are
tuning $m_A \simeq 2m_{LSP}$, allowing neutralinos to annihilate efficiently
through an $s$-channel resonance.

The Neutralino Tuning Problem of the MSSM has been known for some time,
but this observation has not
yet spawned anything like the febrile activity engendered by the
Little Hierarchy Problem. Part of the reason is that it is not
straightforward to address the Neutralino Tuning Problem in a physically
meaningful and unambiguous fashion. 
This is most easily seen by attempting to quantify the problem.

The neutralino relic density, assuming a $\Lambda$CDM cosmological history,
is a physical observable that is not conceptually different from, say,
the selectron mass.
Obviously in the limit that
the WMAP errors shrink to zero, the allowed fraction of the CMSSM parameter
space, or any other space of relevant parameters, will also shrink to zero.
The ``small'' size of the WMAP--allowed region of the parameter space, in and of itself,
therefore has no bearing on naturalness. It is simply a mapping from
the parameter space to a particular physical observable with a small error bar. 
The same is true for more sophisticated statistical 
analyses \cite{Ellis:2003si}-\cite{Allanach:2006jc}.

Another quantitative approach comes from expanding the SUSY parameter space, and comparing
the WMAP--allowed regions for the relic density with those obtained in the CMSSM.
But this approach has serious drawbacks. 
If we expand our class of SUSY models
by varying parameters which are irrelevant to the determination of the
neutralino relic density, we have done nothing.
If we expand our class of SUSY models
by varying relevant parameters, we are also (in general) changing the
measure on ``theory space'', so a naive comparison of allowed regions
is not meaningful. This makes it difficult, for example, to compare
the CMSSM with the non-universal Higgs models defined in \cite{Baer:2005bu},
where the focus point and $A$--funnel regions are complicated slices through
a higher dimensional parameter space.

\begin{figure}
\centerline{\epsfxsize 4.75 truein \epsfbox {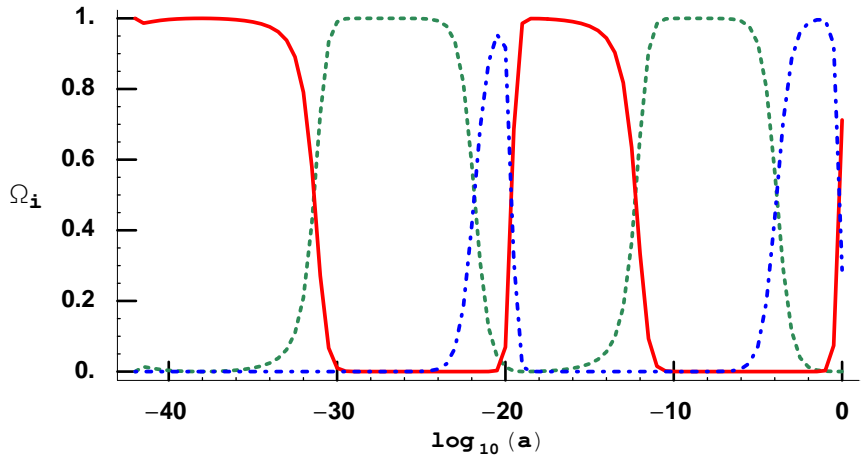}}
\caption{\label{fig:model2} The cosmological history of
Slinky. Shown are the relative energy density fractions
in radiation (green/dashed), matter(blue/dot-dashed), and the noncanonical
scalar (red/solid), as a function of the logarithm of the scale factor $a(t)$.
The early period of (exotic) matter domination is illustrative and has no connection
to neutralinos.
\hbox to170pt{}}
\end{figure}

It may be possible to find a class of SUSY
models which do significantly better than the CMSSM in an unambiguous,
apples--to--apples comparison. However we would not learn much from this unless the
new models had some theoretical or experimental motivation independent from
our desire to explain dark matter.

Experimental motivation could come from the Large Hadron Collider (LHC). 
Initial physics runs may provide
direct access to Terascale supersymmetry. This would certainly give
a dramatic, but not complete, reduction of the SUSY parameter space
compatible with particle physics data. It is possible that in a large fraction
of this reduced parameter space the predicted neutralino relic density
will satisfy the WMAP (or Planck) bounds.  While this would not answer our
original question of whether generic Terascale SUSY gives a natural
explanation of dark matter, it would make this question somewhat moot.
In such a case emphasis will naturally shift to trying to obtain
precision comparable to the WMAP errors, and ultimately to assessing
whether the neutralino constitutes the only major component of dark matter.
It will not be possible to do this using only initial results from the LHC. 
Detailed studies \cite{Baltz:2006fm} indicate that the International Linear Collider
(ILC), combined with the LHC data as well as data from direct and indirect dark matter
searches, will be required to provide adequate resolution 
of the relevant parameters.

In the meantime, it is prudent to assess our original assumptions about
SUSY and about cosmology. Regarding SUSY it is certainly important to consider
LSPs other than the neutralino. However as has been already noted
by many authors \cite{Feng:2005ee},
without simultaneously modifying our
cosmological assumptions this tends to complicate the natural connection
between SUSY and dark matter, rather than simplifying it. 

Regarding cosmology, we have made two major assumptions.
The first is
that neutralinos are the overwhelming dominant component of dark matter.
This is not very plausible, given that visible matter consists of several
quite different stable components. On the other hand this assumption also
has little impact on our analysis of the CMSSM, for the large part of
the parameter space where the LSP is nearly 100\% Bino. Here we generically
predict far too much dark matter. Thus if
we instead asked that the neutralino merely constitute a {\it significant}
fraction of dark matter, our unhappiness with the CMSSM would be unchanged.

\begin{figure}
\centerline{\epsfxsize 3.0 truein \epsfbox {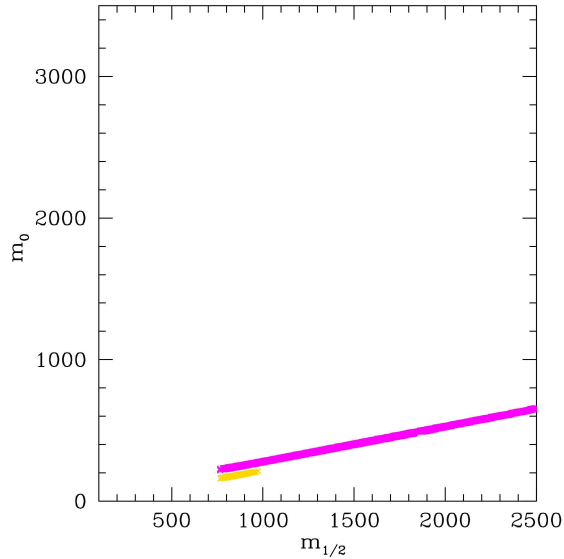}}
\caption{\label{fig:comb5fs} The WMAP--allowed points
for the CMSSM, with $A=0$, $\mu >0$, and tan$\,\beta =5$.
The scan was performed in 2 GeV increments.
\hbox to170pt{}}
\end{figure}

\begin{figure}
\centerline{\epsfxsize 3.0 truein \epsfbox {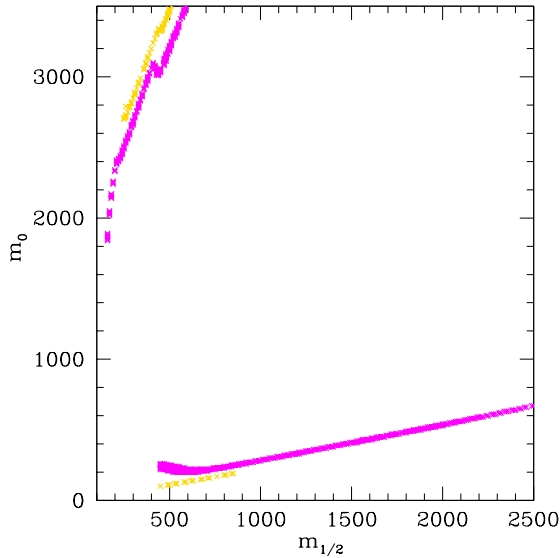}}
\caption{\label{fig:comb10fs} The WMAP--allowed points
for the CMSSM, with $A=0$, $\mu >0$, and tan$\,\beta =10$.
The scan was performed in 10 GeV increments.
\hbox to170pt{}}
\end{figure}

Our second major cosmological assumption was that we can naively
extrapolate the $\Lambda$CDM picture back to the time of neutralino
freeze--out, and that the neutralinos themselves are entirely thermal relics.
This assumption is plausible but not yet directly supported by any data.
The astrophysical data that supports the standard $\Lambda$CDM model
is largely
silent on the cosmological history of the universe before BBN.
WMAP data is consistent with an epoch of primordial inflation, but we are
very far from being able to connect these hints to definitive statements
about dark matter. 

Simply put, a top-down approach of trying to fix pre-BBN cosmology
sufficiently well to pin down the genesis of dark matter is likely to fail in
the foreseeable future. More realistic is to combine a top-down approach
with a bottom-up approach, which uses collider data and dark matter detection
results. Thus, at the same time that we use data to determine the properties
of dark matter, we are simultaneously using data (some of it the same data)
to determine the cosmology of the Terascale epoch.

\section{A new cosmological benchmark}

With the assumption that neutralinos are thermal relics, their
abundance $Y(X)$ can be computed as a function of $X=T/m_{\chi}$,
the ratio of the temperature of the thermal bath divided by the mass
of the LSP. The evolution equation for $Y(X)$ is \cite{Gelmini:1990je,Edsjo:1997bg}:

\begin{eqnarray}
 \frac{dY}{dX}= \frac{m_{\chi}}{X^2}
\sqrt{\frac{\pi  g_*(m_{\chi}/X)) }{45}} M_p <\sigma v>
\left(Y_{eq}(X)^2-Y(X)^2\right)
    \label{eqn:evolution}
\end{eqnarray}
where $g_{*}$ is an effective number of degree of freedom,
$M_p$ is the Planck mass, $Y_{eq}(X)$
the thermal equilibrium abundance, and $<\sigma v>$ is the
relativistic thermally averaged annihilation cross section. The
SUSY model determines the cross section, summing over the relevant
annihilation and co-annihilation channels.

We have used {\tt micrOMEGAs\,2.0}, which solves (\ref{eqn:evolution})
numerically \cite{Belanger:2006is,Belanger:2004yn} 
and returns the predicted neutralino relic density
as $\Omega_{\chi} h^2$. It also returns the freeze--out value
of $X$, called $X_F$, which is roughly equal to 1/20. 
The program uses {\tt SUSPECT\,2.3} to compute
the SUSY spectrum from CMSSM input parameters \cite{Djouadi:2002ze}.
As noted in the previous
section we fix the top quark mass to be 172.7 GeV.

\begin{figure}
\centerline{\epsfxsize 3.0 truein \epsfbox {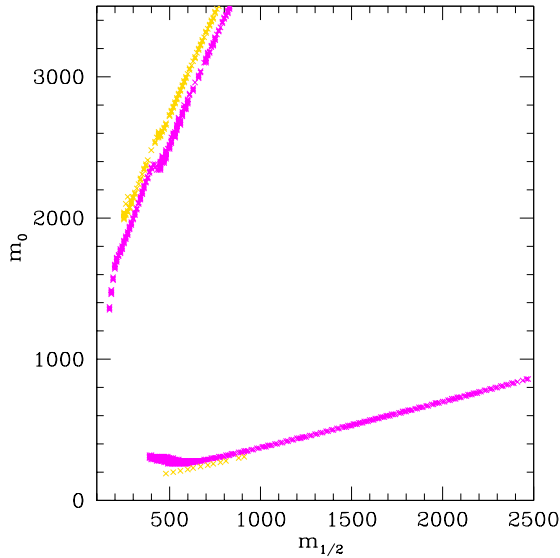}}
\caption{\label{fig:comb30fs} The WMAP--allowed points
for the CMSSM, with $A=0$, $\mu >0$, and tan$\,\beta =30$.
The scan was performed in 10 GeV increments.
\hbox to170pt{}}
\end{figure}

In the Slinky model of quintessential inflation, a single noncanonical inflaton is
responsible for both primordial inflation and the present day accelerating
expansion. The inflaton potential is an exponential of a periodic function,
giving repeated epochs of progressively weaker accelerated expansion, alternating
with epochs of radiation or matter domination. The period is adjusted so that
BBN occurs during a period of radiation domination. The remaining free parameter,
which is the coefficient controlling the coupling of the inflaton
to radiation ($k_0 $), is always small, and is turned off at late times to avoid
strong phenomenological constraints. This weak coupling is fixed by requiring
that the present day ratio of radiation density to quintessence takes the
WMAP--preferred value.
The original Slinky allowed for the nonthermal production of dark matter via
coupling to the inflaton; for the purposes of this paper we assume that this coupling
is not present, and that dark matter is produced thermally entirely from
neutralino freeze--out. 

Thus Slinky is a simple well--motivated cosmological model with no adjustable parameters.
Not surprisingly, at the time that neutralinos freeze out, the Hubble expansion rate
$H$ differs from what it would be in standard $\Lambda$CDM, since the universe is
near the end of an epoch of accelerated expansion. Furthermore, between the time that
the neutralinos freeze out and BBN, there is a significant entropy increase due to
the inflaton coupling to radiation. The net effect is to dilute the neutralino
relic density relative to what would be predicted with $\Lambda$CDM.
Thus $\Omega_{\chi}$(Slinky) $= \Omega_{\chi}$(standard)$/\gamma$, where $\gamma$
is the Slinky dilution factor.
From a numerical analysis, we have found a heuristic formula for $\gamma$
as a function of $X_F$:
\begin{eqnarray}\label{eqn:dilution}
\gamma= 0.6 + 0.5\sqrt{X_F} -0.06 X_F + 0.01 X_F^2 \; .
\end{eqnarray}
This formula is accurate to within a few percent. The exact solution can be
obtained by calculating the entropy production through 
numerically solving the evolution equations for the 
inflaton, radiation and matter, {\it i.e.}
\begin{eqnarray}
\dot{\rho}_{\theta} &=& -3H(1+w)\rho_{\theta} 
-k_0 H (1+w)\rho_{\theta}
\; ;\nonumber\\
\dot{\rho}_r &=& -4H \rho_r 
+k_0 H (1+w)\rho_{\theta}
\; ; \\
\dot{\rho}_m &=& -3H \rho_m 
\; ;\nonumber
\label{eqn:coupled}
\end{eqnarray}
where $H$ is the Hubble rate, $w(t)$ is the inflaton equation of state parameter, and
$\rho_{\theta}$, $\rho_r$, $\rho_m$ are the energy density of the inflaton,
radiation and matter, respectively.

\section{Results}

\begin{figure}
\centerline{\epsfxsize 3.0 truein \epsfbox {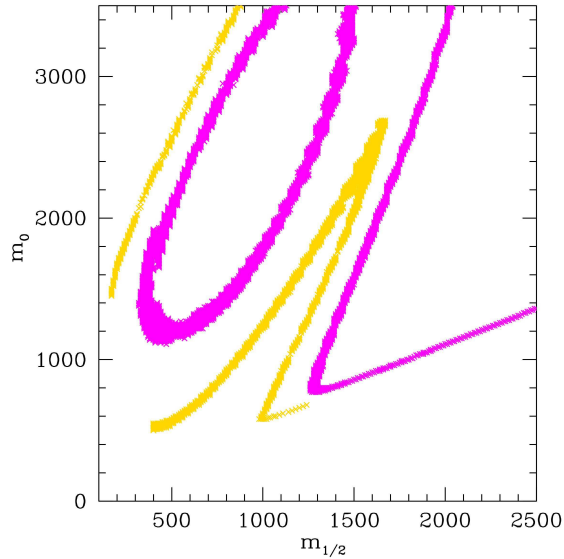}}
\caption{\label{fig:comb50fs} The WMAP--allowed points
for the CMSSM, with $A=0$, $\mu >0$, and tan$\,\beta =50$.
The scan was performed in 10 GeV increments.
\hbox to170pt{}}
\end{figure}

The results of the CMSSM scans are shown in Figures (\ref{fig:comb5fs}-\ref{fig:comb50fs}),
for $A=0$, $\mu >0$, and various values of tan$\,\beta$. We only show points
for which the neutralino is the LSP, we obtain proper electroweak symmetry breaking,
and we satisfy current experimental bounds including $m_h \ge 114$ GeV. The yellow
points predict neutralino relic densities which satisfy the WMAP
bounds (\ref{eqn:wmapbounds}). 
The magenta points predict neutralino relic densities which satisfy the WMAP bounds
when combined with the Slinky dilution factor as given by (\ref{eqn:dilution}).
These points are consistent with all accelerator bounds, and we have checked
using {\tt DarkSUSY\,3.14} \cite{Gondolo:2004sc}
that the resulting spin-independent LSP-nucleon cross sections
are below the CDMS limits \cite{Akerib:2004fq}.

As seen in Table 1, the WMAP--allowed region is an order of magnitude larger
if we assume the modified Slinky cosmology rather than $\Lambda$CDM cosmology.
In the large tan$\,\beta \sim 50$ window this ratio is reduced to less than 3, but
recall that this window is itself a (small) tuning. The large ratio for the case
tan$\,\beta = 5$ is enhanced by the lack of an allowed focus point region for
the standard case.
This comparison of allowed
regions is only weakly dependent on the size of the WMAP error bars, or on the
choice of metric in the ``theory space''.

From a detailed comparison of the particle spectra, we find that the WMAP--allowed
points in the Slinky case are more generic. This is not obvious from the figures, since
except for the case of tan$\,\beta = 50$ the magenta regions are fairly close to
the yellow regions where stau co-annihilation or Bino-Higgsino mixing is occurring.
Looking at the spectra, we find that the lightest stau is at least 40 GeV heavier
than the LSP in the lower magenta regions, while the Bino fraction in the upper
magenta regions never dips below 0.88. The magenta regions do not
involve the kind of balanced mixings, mass degeneracies, or resonance-inducing mass relations
that characterize the yellow regions. Table 2 shows the particle spectra for some
representative magenta points.

\begin{table}[tb]
\centering
\begin{tabular}{|c|c|c|}
\hline\hline
tan$\,\beta$ & number of allowed & number of allowed \\ [-0.5 ex]
 & standard points & Slinky points \\ [0.5 ex]
\hline
5 & 187 & 4812 \\ \hline
10 & 68 & 435 \\ \hline
30 & 108 & 434 \\ \hline
50 & 1423 & 3578 \\[0.5 ex]
\hline\hline
\end{tabular}
\caption{Results of CMSSM scans in the $m_0$-$m_{1/2}$ plane,
with $A=0$, $\mu > 0$, and various values of tan$\,\beta$.
The scans were performed in 10 GeV increments, except for the
case tan$\,\beta =5$, where the scan was in 2 GeV increments. 
The second column
contains the points which give a WMAP--allowed neutralino relic
density, as reported by {\tt micrOMEGAs\,2.0}. The third column
shows the number of WMAP--allowed points when we include the
additional dilution factor of the Slinky benchmark model.}
\end{table}

\begin{table}[tb]
\centering
\begin{tabular}{|c|c|c|c|c|c|c|c|c|}
\hline\hline
$m_0$ & $m_{1/2}$ & tan$\,\beta$ & $m_h$ & LSP & stau1 & chargino1
& gluino & LSP Bino fraction\\ [0.5 ex]
\hline
240 & 810 & 5 & 114.2 & 341 & 382 & 642 & 1792 & 0.996\\ \hline
200 & 520 & 10 & 114.1 & 188 & 274 & 354 & 1065 & 0.990\\ \hline
1840 & 160 & 10 & 115.1 & 61 & 1822 & 109 & 492 & 0.906\\ \hline
1350 & 170 & 30 & 114 & 65 & 1239 & 113 & 500 & 0.887\\[0.5 ex]
\hline\hline
\end{tabular}
\caption{Superpartner and Higgs masses in GeV for selected
CMSSM points which predict the correct amount of neutralino
dark matter using the Slinky benchmark cosmology.
All points have $A=0$ and $\mu > 0$. The particle spectra
are computed using {\tt SUSPECT\,2.3}.}
\end{table}

\section{Conclusion}

In this paper we have reviewed the Neutralino Tuning Problem of the CMSSM and
MSSM. We have quantified this problem by an explicit comparison of the
CMSSM for two different cosmological benchmark scenarios: standard $\Lambda$CDM
and Slinky. The comparison shows that the standard case requires about an order
of magnitude of extra tuning to successfully account for dark matter. The
WMAP--allowed regions for the Slinky case are still restricted to rather
narrow strips, but this largely reflects the small ($\pm$ 10\%) error bar
on $\Omega_{\chi}h^2$ that we are trying to squeeze into. 

We expect that it should be possible to construct interesting
new cosmological benchmarks from examples already in the 
literature \cite{Giudice:2000ex}-\cite{Drees:2006vh}. 
Note that a different cosmological
benchmark with a larger dilution factor would look even more natural than
Slinky. Suppose for example that we fix $m_0$, $A$, sign($\mu$) and tan$\,\beta$,
and look at 
\begin{eqnarray}
f(m_{1/2}) \equiv \vert d\Omega_{\chi}/dm_{1/2}\vert \; .
\end{eqnarray}
We can regard $f(m_{1/2})$ as a rough naturalness measure;
larger $f(m_{1/2})$ in a region where we obtain WMAP-allowed points
means that we will obtain a narrower strip of allowed points.
Generically $f(m_{1/2})$ is largest at both small and large values
of $m_{1/2}$, passing through zero at some intermediate value of $m_{1/2}$
where $\Omega_{\chi}$ is maximized. Thus a cosmological benchmark with a
larger dilution factor will pick up WMAP-allowed points in a region
where $f(m_{1/2})$ is smaller. In addition, when we have a dilution
factor $\gamma$, the relevant naturalness measure is not
$f(m_{1/2})$ but rather $f(m_{1/2})/\gamma$, since it is the latter
quantity which determines the width of the WMAP-allowed region
in the approximation that $\gamma$ is a slowly varying function of $m_{1/2}$.

The superpartner spectra in the standard and Slinky cases differ by
at least tens of GeV. In addition, the Slinky spectra do not exhibit
the mass degeneracies or resonance-inducing mass relations that occur
in the standard case. Thus it should be possible to discriminate
between these spectra using initial data runs from the LHC. This
opportunity deserves detailed study.

Suppose that experiments at the LHC do discover a spectrum of new
particles consistent with the CMSSM, but not consistent with the
WMAP upper bound as derived assuming a neutralino LSP and standard
cosmology. Then we will face the interesting challenge of trying
simultaneously to refine our knowledge of both the underlying Terascale
physics and the underlying Terascale cosmology. Though both may be
complicated, it seems prudent to develop a strategy that begins with
simple benchmarks. The CMSSM+Slinky benchmark described here is a
step in this direction.

\subsection*{Acknowledgments}
We thank  Marela Carena, Scott Dodelson and Will Kinney for 
useful discussions. 
JL acknowledges the support of the Aspen Center for Physics.
GB is grateful for the hospitality of the Fermilab Theory Group 
and for the support from the Spanish MEC under
Contract PR2006-0195 and  FPA 2005/1678.
Fermilab is operated by Universities Research Association 
Inc. under Contract No. DE-AC02-76CHO3000 with the U.S. Department
of Energy.

\newpage



\begin{thebibliography}{99}


\bibitem{Chung:2003fi}  
D.~J.~H.~Chung, L.~L.~Everett, G.~L.~Kane, S.~F.~King, J.~D.~Lykken and L.~T.~Wang,  
Phys.\ Rept.\  {\bf 407}, 1 (2005)  [arXiv:hep-ph/0312378].  



\bibitem{Spergel:2006hy}
  D.~N.~Spergel {\it et al.},
  ``Wilkinson Microwave Anisotropy Probe (WMAP) three year results:
  Implications for cosmology,''
  arXiv:astro-ph/0603449.


\bibitem{Barbieri:1982eh}
  R.~Barbieri, S.~Ferrara and C.~A.~Savoy,
  Phys.\ Lett.\ B {\bf 119}, 343 (1982).

\bibitem{Hall:1983iz}
  L.~J.~Hall, J.~D.~Lykken and S.~Weinberg,
  Phys.\ Rev.\ D {\bf 27}, 2359 (1983).

\bibitem{Nath:1983aw}
  P.~Nath, R.~Arnowitt and A.~H.~Chamseddine,
  Nucl.\ Phys.\ B {\bf 227}, 121 (1983).


\bibitem{Arkani-Hamed:2006mb}  
N.~Arkani-Hamed, A.~Delgado and G.~F.~Giudice,  
Nucl.\ Phys.\ B {\bf 741}, 108 (2006)  [arXiv:hep-ph/0601041].  


\bibitem{Baer:2002gm}
  H.~Baer, C.~Balazs, A.~Belyaev, J.~K.~Mizukoshi, X.~Tata and Y.~Wang,
  JHEP {\bf 0207}, 050 (2002)
  [arXiv:hep-ph/0205325].

\bibitem{Ellis:2003cw}
  J.~R.~Ellis, K.~A.~Olive, Y.~Santoso and V.~C.~Spanos,
  Phys.\ Lett.\ B {\bf 565}, 176 (2003)
  [arXiv:hep-ph/0303043].

\bibitem{Baer:2003yh}
  H.~Baer and C.~Balazs,
  JCAP {\bf 0305}, 006 (2003)
  [arXiv:hep-ph/0303114].

\bibitem{Chattopadhyay:2003xi}
  U.~Chattopadhyay, A.~Corsetti and P.~Nath,
  Phys.\ Rev.\ D {\bf 68}, 035005 (2003)
  [arXiv:hep-ph/0303201].

\bibitem{Battaglia:2003ab}
  M.~Battaglia, A.~De Roeck, J.~R.~Ellis, F.~Gianotti, K.~A.~Olive and L.~Pape,
  Eur.\ Phys.\ J.\ C {\bf 33}, 273 (2004)
  [arXiv:hep-ph/0306219].

\bibitem{Arnowitt:2003vw}
  R.~Arnowitt, B.~Dutta and B.~Hu,
  arXiv:hep-ph/0310103.

\bibitem{Baer:2003ru}
  H.~Baer, A.~Belyaev, T.~Krupovnickas and X.~Tata,
  JHEP {\bf 0402}, 007 (2004)
  [arXiv:hep-ph/0311351].

\bibitem{Gomez:2004ek}
  M.~E.~Gomez, T.~Ibrahim, P.~Nath and S.~Skadhauge,
  Phys.\ Rev.\ D {\bf 70}, 035014 (2004)
  [arXiv:hep-ph/0404025].

\bibitem{Baer:2004qq}
  H.~Baer, A.~Belyaev, T.~Krupovnickas and J.~O'Farrill,
  JCAP {\bf 0408}, 005 (2004)
  [arXiv:hep-ph/0405210].

\bibitem{Ellis:2004bx}
  J.~R.~Ellis, K.~A.~Olive, Y.~Santoso and V.~C.~Spanos,
  Phys.\ Lett.\ B {\bf 603}, 51 (2004)
  [arXiv:hep-ph/0408118].

\bibitem{Allanach:2004xn}
  B.~C.~Allanach, G.~Belanger, F.~Boudjema and A.~Pukhov,
  JHEP {\bf 0412}, 020 (2004)
  [arXiv:hep-ph/0410091].

\bibitem{Mambrini:2005cp}
  Y.~Mambrini and E.~Nezri,
  arXiv:hep-ph/0507263.



\bibitem{Barbieri:2000gf}
  R.~Barbieri and A.~Strumia,
  arXiv:hep-ph/0007265.

\bibitem{Birkedal:2004xi}
  A.~Birkedal, Z.~Chacko and M.~K.~Gaillard,
  JHEP {\bf 0410}, 036 (2004)
  [arXiv:hep-ph/0404197].


\bibitem{Barenboim:2005np}
  G.~Barenboim and J.~D.~Lykken,
  Phys.\ Lett.\ B {\bf 633}, 453 (2006)
  [arXiv:astro-ph/0504090].

\bibitem{Barenboim:2006rx}  
G.~Barenboim and J.~D.~Lykken,  
JHEP {\bf 0607}, 016 (2006)  [arXiv:astro-ph/0604528].  


\bibitem{unknown:2005cc}
    [CDF Collaboration],
  arXiv:hep-ex/0507091.

\bibitem{Group:2006xn}
  T.~E.~W.~Group,
  arXiv:hep-ex/0608032.


\bibitem{Djouadi:2006be}
  A.~Djouadi, M.~Drees and J.~L.~Kneur,
  JHEP {\bf 0603}, 033 (2006)
  [arXiv:hep-ph/0602001].


\bibitem{Feng:2005ee}
  J.~L.~Feng,
  Annals Phys.\  {\bf 315}, 2 (2005).


\bibitem{Ellis:2003si}
  J.~R.~Ellis, K.~A.~Olive, Y.~Santoso and V.~C.~Spanos,
  Phys.\ Rev.\ D {\bf 69}, 095004 (2004)
  [arXiv:hep-ph/0310356].

\bibitem{Allanach:2005kz}
  B.~C.~Allanach and C.~G.~Lester,
  Phys.\ Rev.\ D {\bf 73}, 015013 (2006)
  [arXiv:hep-ph/0507283].

\bibitem{Allanach:2006jc}
  B.~C.~Allanach,
  Phys.\ Lett.\ B {\bf 635}, 123 (2006)
  [arXiv:hep-ph/0601089].



\bibitem{Baer:2005bu}
  H.~Baer, A.~Mustafayev, S.~Profumo, A.~Belyaev and X.~Tata,
  JHEP {\bf 0507}, 065 (2005)
  [arXiv:hep-ph/0504001].


\bibitem{Baltz:2006fm}
  E.~A.~Baltz, M.~Battaglia, M.~E.~Peskin and T.~Wizansky,
  arXiv:hep-ph/0602187.


\bibitem{Gelmini:1990je}
  G.~B.~Gelmini, P.~Gondolo and E.~Roulet,
  Nucl.\ Phys.\ B {\bf 351}, 623 (1991).

\bibitem{Edsjo:1997bg}
  J.~Edsjo and P.~Gondolo,
  Phys.\ Rev.\ D {\bf 56}, 1879 (1997)
  [arXiv:hep-ph/9704361].

\bibitem{Belanger:2006is}
  G.~Belanger, F.~Boudjema, A.~Pukhov and A.~Semenov,
   ``micrOMEGAs2.0: A program to calculate the relic density of dark matter in a
  generic model,''
  arXiv:hep-ph/0607059.

\bibitem{Belanger:2004yn}
  G.~Belanger, F.~Boudjema, A.~Pukhov and A.~Semenov,
  Comput.\ Phys.\ Commun.\  {\bf 174}, 577 (2006)
  [arXiv:hep-ph/0405253].

\bibitem{Djouadi:2002ze}
  A.~Djouadi, J.~L.~Kneur and G.~Moultaka,
   ``SuSpect: A Fortran code for the supersymmetric and Higgs particle spectrum
  in the MSSM,''
  arXiv:hep-ph/0211331.


\bibitem{Gondolo:2004sc}
  P.~Gondolo, J.~Edsjo, P.~Ullio, L.~Bergstrom, M.~Schelke and E.~A.~Baltz,
  JCAP {\bf 0407}, 008 (2004)
  [arXiv:astro-ph/0406204].


\bibitem{Akerib:2004fq}
  D.~S.~Akerib {\it et al.}  [CDMS Collaboration],
  Phys.\ Rev.\ Lett.\  {\bf 93}, 211301 (2004)
  [arXiv:astro-ph/0405033].


\bibitem{Giudice:2000ex}  
G.~F.~Giudice, E.~W.~Kolb and A.~Riotto,   
Phys.\ Rev.\ D {\bf 64}, 023508 (2001) [arXiv:hep-ph/0005123].  

\bibitem{Nihei:2004xv}
  T.~Nihei, N.~Okada and O.~Seto,
  Phys.\ Rev.\ D {\bf 71}, 063535 (2005)
  [arXiv:hep-ph/0409219].

\bibitem{Kohri:2005ru}
  K.~Kohri, M.~Yamaguchi and J.~Yokoyama,
  Phys.\ Rev.\ D {\bf 72}, 083510 (2005)
  [arXiv:hep-ph/0502211].

\bibitem{Pallis:2005hm}
  C.~Pallis,
  JCAP {\bf 0510}, 015 (2005)
  [arXiv:hep-ph/0503080].

\bibitem{Pallis:2005bb}
  C.~Pallis,
  Nucl.\ Phys.\ B {\bf 751}, 129 (2006)
  [arXiv:hep-ph/0510234].

\bibitem{Drees:2006vh}
  M.~Drees, H.~Iminniyaz and M.~Kakizaki,
  Phys.\ Rev.\ D {\bf 73}, 123502 (2006)
  [arXiv:hep-ph/0603165].



\end{thebibliography}
\end{document}